\documentclass[a4paper,10pt]{article}

\usepackage{geometry}
\makeatletter\@addtoreset{chapter}{part}\makeatother%
\usepackage[dvipdfmx]{graphicx}
\usepackage[normal,oneline]{caption2}
\usepackage{url}
\usepackage[affil-it]{authblk}
\usepackage{textgreek}
\usepackage{xcolor}
\usepackage{bm}
\usepackage{appendix}

\begin{document}

\providecommand{\keywords}[1]
{
  \small	
  \textbf{\textit{Keywords---}} #1
}

\title{\bf Parametrization Model Motivated from\\
Physical Processes for Studying the\\
Spread of COVID-19 Epidemic}

\author[]{S.~Maltezos}

\affil[]{National Technical University of Athens\\
Physics Department}

\date{}
\maketitle

\pagenumbering{arabic}

\newcommand{\beq}{\begin{equation}}
\newcommand{\eeq}{\end{equation}}

\newcommand{\ben}{\begin{eqnarray}}
\newcommand{\een}{\end{eqnarray}}

\begin{abstract}
\emph{Abstract.}
The outbreak of the new virus COVID-19, beyond the human health risks and loss, has caused also very serious problems in a wide range of human activities, including the basic and applied scientific research, mainly that concern world wide collaborations. It is desirable to all of us to have the prospect of quickly predicting a turning point in the daily cases curve of the disease. In this work we face the problem of COVID-19 virus disease spreading by aiming mostly to create a reliable mathematical model describing this mechanism for an  isolated society, for cities or even for a whole country. Drawing upon similar mechanisms appearing  in the particle detector’s Physics, we concentrated to the so called, ``semi-gaussian'' function of $n$-degree. This approach can provide some very useful advantages in the data analysis of the daily reported cases of the infected people. Applying this model and fitting to the data, reported until the submission of this work, we have determined, among others, the mean infection time for a citizen in the society under study. We also applied and adopted this model to the reported cases in other countries and we have performed useful comparisons and conclusions.     

\noindent
\end{abstract} \hspace{10pt}

{\keywords{COVID-19, Epidemic, Disease spread, Semi-Gaussian}}

\section{Introduction}

The disease of the new virus COVID-19 became a pandemic infecting almost all countries within only three months from the end of 2019 and finally became a pandemic. A large number of papers have been recently published, not only from the medical side but also from the community of Physical Sciences, that is, Physics, Mathematics and Engineering. In parallel, with the medical efforts for discovering drugs and injections against the COVID-19 virus, it is important to develop an appropriate parametrization model in order to evaluate the public restrictions in a country and as well as to predict the route of the disease in future times. This model, compared with the existing epidemic spread models (e.g. SIR, SEIR) \cite{Peng}, \cite{Li} and others \cite{Liu}, can describe the mechanism of affront of the virus in a social ``system'', assumed well isolated or quarantined, and at the same time, the subsequent infection of the virus among the people. The number of the infected people as a function of time can be expressed by a proposed three-parameter mathematical function which can be fitted to the real data by a non-linear optimization process. This model has been motivated from our recent work on particle detector quality evaluation \cite{ATL_Note} but a similar approach has been also proposed and used in \cite{Vazquez} and \cite{Ziff} for the data of other countries or provinces which appeared spreading of the disease earlier. The analyzed data concern the daily reported cases in a country, regardless if that constitutes a small fraction of the actual - non certified cases reported until one day before the submission of this work ($6^{\mathrm{th}}$ of April 2020). Therefore, without commenting on its competitiveness and its effectiveness, it is clear that apart from the assumptions under which it is expressed, it can be a flexible configuration model easily adaptable to a wide variety of geographical areas or cities and can offer early knowledge to the responsible official committee of a country.

\section{Infectiousness mechanism}

The subject of the present work is the new COVID-19 which caused a large number of serious health problems because of its special characteristics. The most underlined one is the very high infectiousness leading to very rapid spreading over the world. The standard epidemiological methods faced to the problem, not only studying its evolution in time but also giving efforts to find satisfactory reduction and defence strategies. In parallel, a mathematical parametrization model is very useful investigating the main qualitative parameters like, for instance, the reproductive number, the degree of growth, the mean infection time and other figures-of merit. Regardless of the parametrization model used, we must pay attention to the collected samples of the reported cases and the achieved the isolation of the country or providence under study. The approach to simplify the parametrization model shows some advantages for understanding the overall dynamical mechanism of the disease spreading, even if due to the limitation of detection methods and diagnostic criteria, the asymptomatic or mild patients are usually excluded from the reported (verified) cases. 

For obtaining reliable results and useful predictions, in our work we demand to maintain the condition of low interconnections between susceptible and infected people in a bordered region. If a number of infected by a virus enter an isolated region or country, the infectiousness starts within a certain time period, let $\tau$, resulting to a multiple infection expressed by the so-called ``average reproductive number'' expressing the dynamics of the spreading mechanism, $R_o$. If $R_o\leq1$ then the average number of new cases decreases exponentially (creating an endemic), while if $R_o>1$ an exponentially happens (creating an epidemic). Therefore, an accurate parametrization (modelling) of the spreading mechanism of the particular disease is of significant importance. 

Many statistical models have been proposed and used, while the mathematical parametrization models based on dynamical systems are less usable and generally are given less attention. However, they can offer and provide essential tools for understanding better the mechanisms. The classical Susceptible Exposed Infectious Recovered model (SEIR) with its notable generalizations is the most widely used with high effectiveness of various measures. 

\section{The semi-gaussian parametrization model}
\label{SG}
The proposed model is based on the exponential distribution of the probability theory and statistics. This distribution, with the geometrical one, are the only unmemorable distributions. The events of illness in an epidemic happen randomly and their statistical distribution has the property of ``lack of memory''. Moreover, the mean rate of the events is considered to be constant related to the characteristics of the particular disease (mean infection time) in convolution with the recovering time. Because the ``measurable'' quantity in practice for this issue is the daily counting of cases, the corresponding model must be a function of time. What we propose is the so called $n-$degree ``semi-gaussian'' (SG). This function is applied also in electronics describing the response of a CR-(RC)$^n$ shaper to a unit step signal input and shows some interesting properties. This function is

\beq
c(t)=\frac{1}{n!}\left(\frac{t}{\tau}\right)^{n}{e}^{-t/\tau}
\label{rc_cr}
\eeq

Based on this function, a parametrization model for describing the daily reported cases of an epidemic disease process, let $c(t)$, can be written  

\beq
c(t)=A{{t}^{n}}{{e}^{-t/\tau }}
\label{general}
\eeq

where $A$ is an arbitrary amplitude, $\tau$ is the time constant representing in our case the ``mean infection time'' and $n$ is the degree of the model. It is easy to prove that the ``peaking time'', $t_\mathrm{p}$, depends only on $\tau$, that is, $t_\mathrm{p}=n\tau$.

The maximum value of $c(t)$ is given by

\beq
{{c}_{\max}}=A{{\left( n\tau \right)}^{n}}{{e}^{-n\tau /\tau }}=A\left(\frac{n\tau}{e}\right)^n  
\eeq

In the special case where $n$ is natural number, the following interesting properties can be observed: using the general identity, 
$\int\limits_{0}^{+\infty }{{{t}^{n}}{{e}^{-at}}\mathrm{dt}}=n!/{{a}^{n+1}}$, with $n\geq 1$ and $a>0$ applied to Eq. \ref{general}, we obtain

\beq
C=An!{{\tau }^{n+1}}=A\Gamma{(n+1)}{{\tau }^{n+1}} 
\label{integration}
\eeq

Moreover, let us calculate the integral to peak ratio, $f$.

\beq
f=\frac{C}{{{c}_{\max }}}=\frac{\Gamma{(n+1)}{{e}^{n}}}{{{n}^{n}}}\tau 
\label{ratio}
\eeq

we observe that, for a given $n$ the ratio $f$ depends only on $\tau$ in a linear way, and therefore, the higher the $\tau$ the higher the ratio $f$. This is important conclusion in the strategy for facing the disease spreading using the method of isolation of the people (quarantine). Certainly, the model cannot perfectly describe the epidemic without some particular assumptions and constant parameters, as we can see in the next. The SG function, due to its mathematical property is not symmetrical, but for $n>3$ its shape tends to be symmetrical, a property which is useful feature in the fitting to the data in our application.

By another view, according to the statistical analysis done in \cite{Vazquez}, the spreading mechanism can be modelled as a branching process, by which the obtained function takes the shape a $n-$degree polynomial growth followed by an exponential decay. 

\beq
c(t)\approx \frac{RR_o^n}{\tau n!}\left(\frac{t}{\tau}\right)^{n}{e}^{-t/\tau}
\label{vazquez}
\eeq

where $R_o$ is the reproductive number and $R$ its initial value, affecting also the size of the network in the referred paper and assumed constant. This function is similar with that given in Eq. \ref{rc_cr}, apart of the existence of the additional parameter of $R_o$ in $n^{\mathrm{th}}$ power. Under the hypothesis that $n$ is a natural number and the associated assumptions, this function can be also applied to estimate, among the other free parameters, the reproductive number $R_o$ by fitting to the raw data.

In the application under study of spread of the disease, we must fit the data of the daily reported cases (DRC) of the disease and in the subsequent phase, using the derived three parameters, to calculate the running integral in comparison with the cumulative sum of the data corresponding to the total reported cases (TRC). There are two methodologies to be applied, as follows:

\begin{enumerate}
\item{Fitting the model to the DRC raw data, and then, calculating the running integral.}
\item{Fitting the integral of the SG model to TRC raw data, and then, calculating inversely the SG model itself}
\end{enumerate}  

According to the first methodology, the data are fitted with the function given in Eq. \ref{general}, considering $n$ as a free parameter (not as natural number) for having one more degree of freedom. We use the following free parameters, $p_1=A$, $p_2=n$ and $p_3=\tau$.

\beq
C(t)=p_1 t^{p_2}e^{-t/p_3}
\label{first_method}
\eeq

While the first methodology is definitely the classical one, the observed high fluctuations of the raw data of the DRC lead us to introduce the second methodology. It is reasonable that due to the nature of the integration procedure the fluctuations of the TRC raw data are much lower. The running integral of the SG model, $C(t)$, is calculated as following

\beq
C(t)=\int\limits_{0}^{t}{C(x)\mathrm{d}x}=\int\limits_{0}^{t}{A{{x}^{n}}{{e}^{-x/\tau }}\mathrm{d}x}=A{{\tau }^{n}}\int\limits_{0}^{t}{{\left(\frac{x}{\tau } \right)}^{n}}{{e}^{-x/\tau }}\mathrm{d}x=A{{\tau }^{n+1}}\int\limits_{0}^{t/\tau }{{{u}^{\left( n+1 \right)-1}}{e}^{-u}}\mathrm{du}
\eeq

But we know that

\beq
\int\limits_{0}^{t/\tau }{{{u}^{\left( n+1 \right)-1}}{{e}^{-u}}\mathrm{d}u}= \Gamma (n+1)-\Gamma_\mathrm{c}(n+1,t/\tau) 
\eeq

where the symbol $\Gamma_\mathrm{c}$ represents the upper incomplete gamma function. Upon the above identity we can obtain the expression of the running integral and the corresponding fitting model

\beq
C(t)=A{{\tau }^{1+n}}\left[ \Gamma (n+1)-\Gamma_\mathrm{c}(n+1,t/\tau)\right] 
\eeq

\beq
C(t)=p_1p_3^{1+p_2}\left[ \Gamma (p_2+1)-\Gamma_\mathrm{c}(p_2+1,t/p_3)\right] 
\label{second_method}
\eeq

An important fact concerning the fitted parameters, is the sensitivity to the overall integral given in Eq. \ref{integration} for $n$ natural number. The ratio of the sensitivities of the parameters $n$ and $T$ for very small, variations has been calculated as follows

\beq
\frac{\partial C}{\partial n}\frac{{{n}_{o}}}{{{C}_{o}}}/\frac{\partial C}{\partial \tau }\frac{{{\tau }_{o}}}{{{C}_{o}}}=17 
\label{sensitivity}
\eeq

where, $C_o$ is the overall integral and $n_o$ and $\tau_o$ are the nominal values of the parameter at present time $t_o$ for the data of Greece. Therefore, the model is 17 times more sensitive to the parameter $n$ compared to the parameter $\tau$, and thus, we must pay our attention to the role of it to the spreading mechanism.

We also can calculate the peak-to-integral ratio (PIR) in the general case where $n$ is a real number, as $p=\frac{c_{max}}{C_o}$ in units of $\mathrm{days^{-1}}$. This is a very useful figure-of-merit for evaluating the degree of flattening the SG (the smaller the $p$ the better the flatness. This, essentially, expresses the overshot level with respect to the average level of the SG curve, which is the reverse of the so called in the literature, ``equivalent width'' .

\begin{figure}[!ht]
\centering
\begin{minipage}[b]{0.3\linewidth}
\centering
\includegraphics[scale=0.35]{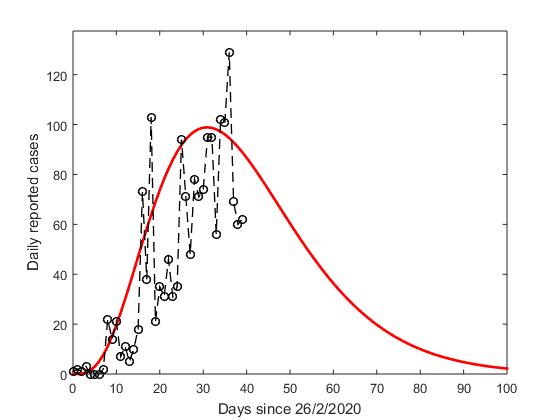}  
\caption[]{Plot of daily reported cases in Greece (black circles) and the optimal curve of the fitted model (red solid line).}
\label{daily}
\end{minipage}
\qquad
\begin{minipage}[b]{0.51\linewidth}
\centering
\includegraphics[scale=0.35]{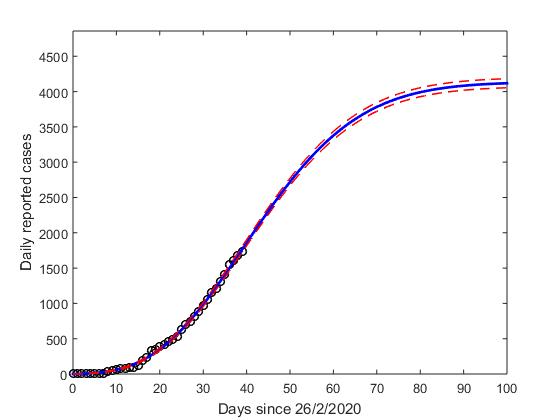}
\caption[]{Plot of the total reported cases in Greece (black circles) and the optimal cumulative curve based on the proposed model (blue solid line). The uncertainty zone of $1\sigma$ is delimited among the two red dashed lines.}
\label{total}
\end{minipage}
\end{figure}

\section{Application to the data in various countries}

\subsection{Results for Greece}

In Greece, during the first month passed from the first reported case (26/2/2020) \cite{worldometers}, we observed that the reported data were appearing with increased statistical fluctuations of the order of $\pm30\%$ (peak-to-valley within a few days). This was continued until present time (April 2020). This likely is due partially to the low statistics and even to other unknown parameters related to the location of the bulk of the cases, if it was in Athens or in other cities.

\subsection{Results for other countries}

\begin{figure}[!ht]
\centering
\begin{minipage}[b]{0.3\linewidth}
\centering
\includegraphics[scale=0.34]{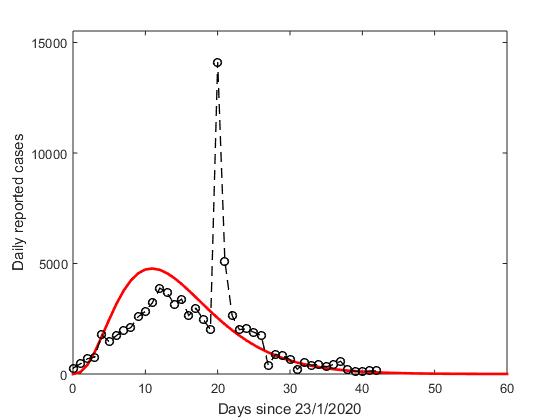} 
\caption[]{Plot of daily reported cases in China (black circles) and the optimal curve of the fitted model (red solid line).}
\label{daily_china}
\end{minipage}
\qquad
\begin{minipage}[b]{0.50\linewidth}
\centering
\includegraphics[scale=0.34]{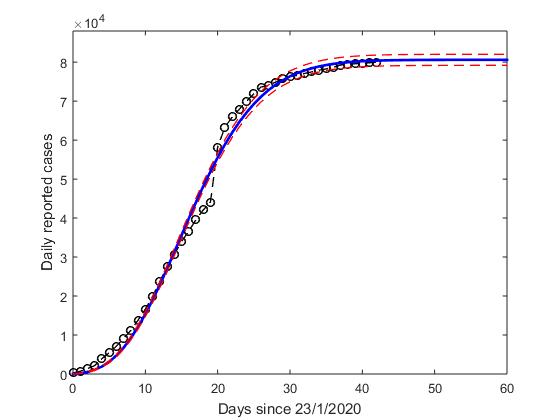}
\caption[]{Plot of the total reported cases in China (black circles) and the optimal cumulative curve based on the proposed model (blue solid line). The uncertainty zone of $5\sigma$ is delimited among the two red dashed lines.}
\label{total_china}
\end{minipage}
\end{figure}

\begin{figure}[!ht]
\centering
\begin{minipage}[b]{0.3\linewidth}
\centering
\includegraphics[scale=0.35]{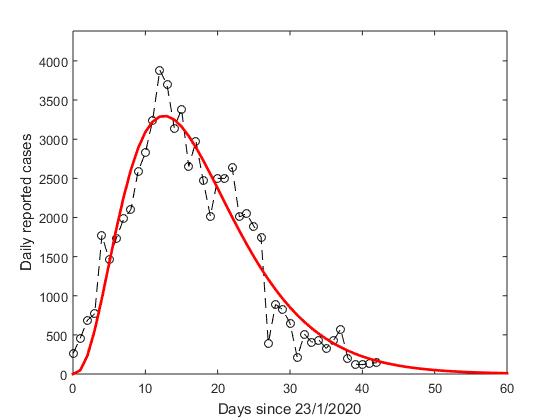} 
\caption[]{Plot of the modified daily reported cases in China (black circles) and the optimal curve of the fitted model (red solid line).}
\label{daily_china_mod}
\end{minipage}
\qquad
\begin{minipage}[b]{0.45\linewidth}
\centering
\includegraphics[scale=0.35]{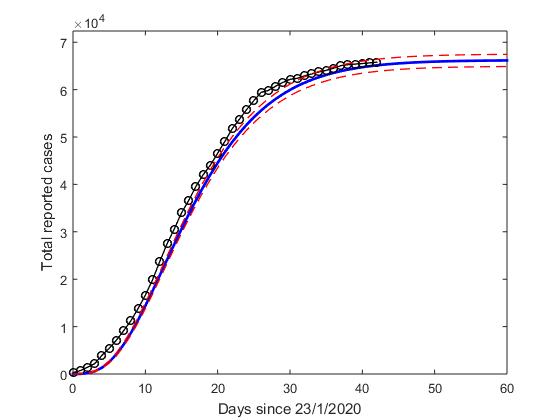}
\caption[]{Plot of the modified total reported cases in China (black circles) and the optimal cumulative curve based on the proposed model (blue solid line). The uncertainty zone of $5\sigma$ is delimited among the two red dashed lines.}
\label{total_china_mod}
\end{minipage}
\end{figure}

By applying the model to the data of China, where the first case was reported by 23th of January, we obtained the results shown in Fig.  \ref{daily_china}and Fig. \ref{total}. The peak in the daily reported cases on 12 of February 2020 is caused to the change in the strategy to start the registration of the patients in province of Hubei certified also by chest X-ray radiography. But this has happened temporarily. Therefore, the analysis must be done also without this peak for obtaining more reliable results, consistent with the data until this day. We replaced the reported number by the average of the two neighbouring values. The obtained results are presented in Fig. \ref{daily_china_mod} and Fig. \ref{total_china_mod}.

\begin{table}[!h]
\centering
\begin{tabular}{|c|c|c|c|c|c|c|}
\hline 
COUNTRY & STAGE & $t_d$ [days]&  $n$ & $\tau$ [days] & $t_p$ [days] \\ 
\hline 
Greece & Turning point &  - &  3.6 & 8.6 & 31  \\ 
\hline 
China & Elimination tail & - & 1.5 & 5.3 & 13  \\ 
\hline 
South Korea & Elimination tail & - & 3.6 & 2.3 & 8.3  \\ 
\hline 
Italy & Turning point & - & 6.8 & 4.7 & 32  \\  
\hline
Switzerland & Turning point & - & 5.2 & 4.4 & 23  \\  
\hline
United States & Growing & 7 & - & - & -  \\ 
\hline
United Kingdom & Growing & 7 & - & - & -  \\ 
\hline 
\end{tabular} 
\caption{Summary of the results obtained by using the proposed model. The turning point corresponds to the peak of the daily reported cases. The peaking time has been estimated by the model.} 
\label{results}
\end{table}

\subsection{Investigation and evaluation of the results}

Studying the obtained results, we can analyse the parameters one-by-one to derive useful conclusions relevant the size of the country, mentality, discipline and defence possibilities of their citizens. Let us start with the parameter $A$. This, for sure, is a scaling parameter and ``measures'' the size (population) of the country, and apart of its severity for the inconvenience of a large number of people, it doesn't give information on the mechanism of the disease.
The exponent, $n$, expresses the rising rate of the DRC curve which is translated to how fast the citizens were moved to distant places or to other cities during the first stages of the disease. 
The mean infection time, $\tau$, expresses the overall achieved isolation among the citizens in the cities and between the cities in a country without focusing to the details of the mechanism.

Another important aspect is the sensitivity of the above parameters. According to the result given by Eq. \ref{sensitivity} the parameter $n$ is extremely crucial trying to maintain the total cases in low levels. Observing the values presented in Table \ref{results} we can identify the higher values in Italy and Switzerland which can lead to greater total cases at the elimination stage of the disease.  
In order to compare and mainly to correlate the obtained parameters among the various countries, it is necessary first to discriminate and categorize them by means of the spreading stage that they are in the present time.
A reasonable and practical criterion to identify the present spreading stage, is the so called, ``time for doubling cases'', let $t_d$. This cannot be expressed mathematically by a simple - practical formula but can be calculated numerically. The condition, $t_d^c\leq n\tau$ seems to be reasonable because the $\tau$ ``measures'' the peak position at DRC curve, $n\tau$ approximately (it holds in the case $n$ being natural number) and there is not any possibility to predict it in early stage. Therefore, if $t_d<t_p/2=n\tau/2$, that is, at half time before the peak, the $t_d$ should be the only useful determined parameter by the optimization and is calculated in the present time $t$(see Table\ref{results}).

According to this criterion, the two categories are: a) that with $t_\mathrm{d}>t_\mathrm{p}^\mathrm{c}/2$, including Greece, Italy, Switzerland, China and South Korea, and b) with $t_\mathrm{d}<t_\mathrm{d}^\mathrm{c}/2$, United States ($t_\mathrm{d}=6$ days) and United Kingdom ($t_\mathrm{d}=6$ days). In category (a) the three model parameters are reliable, while in (a) only the rising branch of the curve can be parametrized and provided together with $t_\mathrm{d}$.

\begin{figure}[h!]
\centering
\includegraphics[width=10cm]{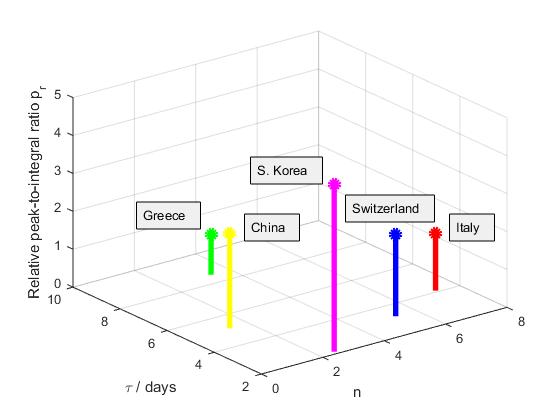}
\caption[]{Plot of the ratio $p_\mathrm{r}$ in the parametric space. For Greece (green) the value is equal to 1 (reference), while for China is 2.43 (yellow), for S. Korea 4.35 (magenta), for Italy 1.43 (red) and for Switzerland 2.08 (blue). The low $p_\mathrm{r}$, the better in controlling the spreading rate of the disease.}
\label{pir}
\end{figure}

Beyond the previous investigation, we use the dimensionless figure-of-merit based on definition of PIR in section \ref{SG}, according to Eq. \ref{ratio}, as relative quantity, $p_\mathrm{r,i}=p_{\mathrm{i}}/p_{\mathrm{1}}$, where the index $1$ refers to Greece in our case. This number, referred to one particular country, expresses the flattening of the DRC curve (or peak suppress) related to the degree of success of the public policy in conjunction with other aspects relevant to the country and its urban or geographic topography. The obtained numbers are presented in the parametric space of $n$ and $\tau$, as shown in Fig. \ref{pir}.
From mathematical point of view, the higher $n$ the lower $p_\mathrm{r}$ with half rate (a $+10\%$ variation causes about $-5\%$ in $p_\mathrm{r}$), and as well as, the higher $\tau$ the lower $p_\mathrm{r}$ (a $+10\%$ variation causes about $-15\%$ in $p_\mathrm{r}$), both calculated around the average of range of the fitted values. 

In our study, the $n$-direction expresses the citizen's massive transportation or local circulation in a country while the $\tau$-direction the isolation degree of them. We can identify the two characteristic cases, the Italy with high value of $n$ (showing the massive people transportation during the epidemic) and the Greece with high value of $\tau$ (indeed with success in this behaviour). Also, in Greece the $p_\mathrm{r}$ value is smaller among the countries under study (all have values greater than the unit which the reference one). Certainly, considering these two parameters, the best result can be accomplished by compromising the desired figures-of-merit, depending on the policy of each country.
However, in Greece, the situation can be characterized very good, by means of the absolute numbers of cases and deaths and as well as of the flattening of the DSG curve and that the health system is under control compared to the other countries. It is hard to explain this success in details. There are, not only social, demographical, geographical and topographical (including many islands) differences but also in public policy, like the early applied social rules and restrictions. Moreover, we must take into account the level of traffic connections and other unknown parameters.

\section*{Conclusions and prospects}

An parametrization model, motivated from physical and describing the new virus COVID-19 disease growth is presented in this work. This model has the important advantage of easy adaptation to the reported real data of the cases including only three free parameters. By using this model we have analyzed the daily reported data in Greece, and as well as, in other indicative countries, aiming not only to predict the time of reaching the turning point, but also to investigate the behaviour of the associated parameters. In this frame, we have studied systematically the role of the parameters in the real social system presented also an overall figure-of-merit evaluating the flattening of the DSG curve. A first conclusion is that, any accurate prediction is impossible having data only before the turning point of the epidemic, given also the dynamical character of the effect. However, having data near or just after the turning point, a good prediction of the spreading can be done. An overall clear picture can be demonstrated after about two weeks from the first reported case. A second conclusion concerns the way to evaluate the overall spreading of the disease. Apart of how we could regulate the associated fitted parameters by appropriate correctional actions in the public advices or restrictions, with must compromise the flattening of the daily reported cases (or the corresponding curves of inpatients) in expense of the overall infected people. Because we are speaking for human lives this, indeed, very hard to balance. After the elimination of the main outbreak of the epidemic in the most of the countries, we plan to re-analysing the complete data deriving final parametrized model concluding about some basic principles with higher accuracy.   

\noindent

\section*{Acknowledgements}

We would like to thank my colleagues, Prof. T. Alexopoulos and Prof. Emeritus E. Fokitis for their insightful comments and useful discussions.

\end{document}